\documentclass[12point]{article}
\usepackage{graphicx}
\usepackage{amsmath}
\usepackage{setspace}
\title{Emergence of cosmic space and its connection with thermodynamic principles} 
\author{P. B. Krishna $^1$, Hassan Basari V. T.$^1$,and Titus K Mathew$^{1,2}$  \\ 
$^{1}$Department of Physics, Cochin University of Science and Technology, Kochi-22, India.\\
$^2$ Inter University Centre for Studies on Kerala Legacy of  Astronomy and Mathematics, \\ CUSAT, Kochi-22, Kerala, India.\\
 krishnapb@cusat.ac.in; basari@cusat.ac.in;  titus@cusat.ac.in}
\date{}
\begin{document}

\maketitle

\begin{abstract}
The recent research on the connection between gravity and thermodynamics suggests that gravity could be an emergent phenomenon. Following this, Padmanabhan proposed a novel idea that the expansion of the universe can be interpreted as equivalent to the emergence of space with the progress of cosmic time. In this approach, the expansion of the universe is described by what is known as the law of emergence, which states that the expansion of the universe is driven by the difference between the number of bulk and surface degrees of freedom in a region bounded by the Hubble radius. This principle correctly reproduces the standard evolution of a Friedmann universe. We establish the connection of the law of emergence, which is conceptually different from the conventional paradigm to describe cosmology, with other well-established results in thermodynamics. It has been shown that the law of emergence can be derived from the unified first law of thermodynamics, which can then be considered as the backbone of the law. However, the law of emergence is rich in structure than implied by the First law thermodynamics alone. It further explains the evolution of the universe towards a state of maximum horizon entropy. Following this, it can be considered that the first law of thermodynamics, along with the additional constraints imposed by the maximisation of the horizon entropy, can together lead to the law of emergence. In the present article, we first make a brief review of Padmanabhan's proposal and then studies its connection with the thermodynamics of the horizon in the context of Einstein's, Gauss-Bonnet, and more general Lovelock gravity theories.
\end{abstract}

 \section{Introduction}
  General relativity is the most successful theory of gravity, which could explain a variety of observations, including the bending of light, perihelion shift of Mercury and the recent detection of the gravitational waves.
   Despite this great success, there remain some issues with the general relativity  which are not yet resolved \cite{209,232,162,175}. One of the important issues is the presence of singularities \cite{209,162}. It is generally believed that this issue could be resolved by formulating a quantum theory of gravity. 
    But, Since general relativity and quantum mechanics are different at the fundamental level, formulation of a quantum general relativity is a difficult task \cite{175}. Meanwhile, the deep connection between gravitational dynamics
 and thermodynamics motivates the emergent interpretation
 of gravity. Such a connection was realized after the
 discovery of black hole thermodynamics by Bekenstein
 and Hawking \cite{26,Bekenstein1,Hawking1,Hawking2,Bekenstein2,Bekenstein3}. A significant step in this field was put
 forward by Jacobson \cite{Jacob1}. He obtained Einstein's field equations
 from the fundamental Clausius relation on a local
 Rindler causal horizon. Following this, various schemes
 for relating gravity and thermodynamics were discussed
 in a variety of gravity theories \cite{paddystatic,sumanta1,sumanta2,paddy5}. This connection  of gravity with thermodynamics, motivates to consider gravity as the thermodynamics of spacetime. But thermodynamics, is an emergent phenomenon, which deals with the connection of macroscopic variables like, pressure, temperature, volume, etc., which doesn't have any existence in the microscopic world. This in turn implies that, gravity also could be an emergent phenomenon.
 
  In the above idea of emergent gravity paradigm, the equations of gravity are treated as emergent, while the background spacetime is pre-existing. Recently, a novel perspective 
 was suggested by Padmanabhan \cite{Paddy1}, assuming that space-time itself as an emergent structure. Conceptually, it is hard to imagine time as being emerged from some pre geometric variables. It is also difficult to treat the space around finite gravitating systems as emergent. However, these conceptual complexities will disappear in the cosmological context when one chooses the time variable as the proper time of the geodesic observer to whom CMBR appears homogeneous and isotropic. Thus, the emergence of spacetime provides an elegant way of describing the cosmological evolution as the emergence of cosmic space with the progress of cosmic time. He could also arrive at the Friedmann equation for a spatially flat universe from this new idea, in the context of Einstein's gravity. This law of emergence proposed in the context of general relativity has been generalized by Cai and Yang et. al. to Gauss-Bonnet and Lovelock gravity for a spatially flat universe \cite{Cai1,Yang}. While Cai modified the degrees of freedom and the volume increase in order to arrive at the Friedman equation in each gravity theory, Yang et. al. assumed a more general dynamical equation for describing the rate of emergence without modifying the degrees of freedom. Following this, Padmanabhan's proposal was also extended to a non-flat universe. In \cite{sheykhi}, Sheykhi considered the areal volume for describing the rate of emergence, and the authors of \cite{modthanu} employed proper invariant volume in formulating the law of emergence. It is to be noted that, while Sheykhi could easily arrive at the Friedmann equations, the authors of \cite{modthanu} have to redefine the Planck length as a function of cosmic time in order to reach the Friedmann equations. The modification of a fundamental constant itself is troublesome. Moreover, the modified Planck length seems to diverge in the matter and radiation dominated era \cite{modthanu}.

 According to Padmanabhan's proposal, the evolution of the universe can be interpreted as the emergence of cosmic space with the progress of cosmic time through a simple equation,
 which is dubbed as the law of emergence \cite{Paddy1,Paddy11,Paddybook}. In this paper, we analyze whether this law of emergence has any direct link with the horizon thermodynamics. From a thermodynamic point of view, we shall
 see what the law of emergence implies and what its origin is. We have explored the direct connection between the law of emergence and the horizon thermodynamics. A rapid overview of the changing perspectives on gravity and spacetime is provided in the upcoming section. This section motivates the emergent perspective on gravity and spacetime and provide the necessary background which is essential for the latter parts of this paper. In section 3, we derived the law of emergence from the unified first law of thermodynamics in the context of Einstein, Gauss-Bonnet and Lovelock gravities. Following this, we investigate the feasibility of formulating the law of emergence and the first law of thermodynamics in a non-flat universe in section 4.  In section 5, we explore the connection between the holographic equipartition and the horizon entropy maximization.  We have then engaged in extending these results to Gauss-Bonnet and more general Lovelock gravities for a spatially flat universe. Section 6 presents summary and conclusions.   
 \section{Towards an alternate perspective on gravity and spacetime}
 The first theme of this paper is to provide a broad overview on the changing perspectives on gravity and spacetime during the last century, which forms the necessary background for the rest part of the review. We start with the troubles with classical gravity which is followed by the discussions on the attempts to quantize gravity and the higher dimensional gravity theories. We  then discuss about the connection between gravity and thermodynamics, the notion of an emergent phenomenon and the main claims of the emergent interpretation of gravity. Following this, we describe the evolution of the universe as the emergence of cosmic space.
 \subsection{Troubles with classical gravity}
 Despite its great success and theoretical beauty, the general theory of relativity has some fundamental flaws. Generally,
 the completeness and faultlessness of a theory rely on its predictability and singularity free description. The main
 trouble with general relativity is its incapability in predicting the future evolution of the dynamic variables due to
 the occurrence of singularities in some well specified circumstances (see\cite{209}).  
 
 Classical general relativity is known to have at least two kinds of singularities. One is the black hole singularity, and
 the other is the big bang singularity. When one asks the question,\cite{175}: ``A neutron star of mass 6M$_{\odot}$ collapses to form
 a blackhole. How will the physical phenomena appear with respect to
 a hypothetical observer on the surface of the neutron star at arbitrarily
 late times as measured by the observer's clock?''
 The theory of relativity can not provide
 an acceptable answer to this question. Within the premises of general relativity, one can not say what happens
 to the observer who arrives at the singularity in a finite proper time. Apart from this, classical general
 relativity can not say anything about the initial stage of the universe that had existed 20 billion years ago. When
 we integrate the Friedmann equation backwards, it predicts the existence of big bang singularity within the cosmic time
 $t<20$ billion years. The existence of these singularities makes general relativity incomplete and thus questionable\cite{162}.   
 It is also worth mentioning that some unique features of Einstein's gravity lead to unexpected difficulties
 when one attempts to obtain field equations from the Einstein-Hilbert action\cite{232}.

 \subsection{Attempts to quantize gravity}
 
 We have discussed about the breakdown of general relativity at the singularities in the last section. There is a
 prevailing consensus that these singularities could be removed by combining general relativity with the other pillar
 of modern physics, the quantum mechanics. It is worth mentioning that general relativity and quantum mechanics could
 separately explain a large number of experiments accurately. However, these two theories seem to be inconsistent with each
 other since they are different at the fundamental level\cite{175}. On the other hand, physicists are trying to formulate grand
 unified theories within the premises of quantum field theory. Even though the quantum field theory succeeded in
 unifying the other three fundamental interactions, strong, weak and electromagnetic, it failed to incorporate gravity.
 It also needs to be mentioned that some physicists are not satisfied with the quantum field theory. While
 Einstein was not comfortable with the conceptual structure of the quantum theory, Dirac criticized the re-normalization
 procedure involved in quantum field theory. 
 
  Also, there arise conceptual issues with quantum theory
 	when one tries to explain the evaporation of black holes. One of the
 	main issues in the black hole context is the black hole information
 	loss paradox, which can be summarised as follows. It is justified to
 	assume that the complete information about the initial state that
 	collapses to form a black hole is either destroyed or encoded in
 	the resulting radiation when the black hole evaporates. However, since
 	the remnant radiation in this process is dominantly thermal, it is
 	impossible to know much about the initial data, although the black
 	hole evaporates completely. This threatens the unitary evolution,
 	the most fundamental feature of quantum theory. Although
 	there can be sources of distortions that can be non-thermal, they are
 	not enough to make the evolution unitary. ( \textit{But, following some aspects
 	of the emergent paradigm, one can expect that these distortions carry some part
 	of the information about the initial state. In reference\cite{Chakraborty2017}, it is shown that a particular kind of distortion can be used to  completely reconstruct a subspace
 	of initial data. It is also argued that one should not ignore the fact that the
 	collapsing material was inherently quantum mechanical in nature. For more
 	details, see reference \cite{Chakraborty2017}.} )

  One can construct a quantum theory of gravity by using the perturbative quantum field theory methods. Even though this program was successful in quantizing the other three interactions, it doesn't work with gravity since gravity is not re-normalizable. String theory\cite{104} and loop quantum gravity \cite{16,213,92} are the two major
 candidate theories of quantum gravity, which are still surviving but not completely successful.

 The quantization of gravity still remains the biggest open problem in fundamental physics. Since it is a challenging problem, it is natural to ask the question,’ Whether the quantization of gravity is the only way to overcome
 the troubles of general relativity?’ Following this, some physicists have attempted to remove the singularities
 by some non-quantum modification or by generalizing Einstein's gravity.  
 
 \subsection{Gravity in higher dimensions}
 Generally, the existence of higher spacial dimensions has been postulated in physical theories for utilizing the extra features of such models. 
 Many of the  physicists support the existence of higher dimensions since they expect to find answers to some questions that can not be 
 answered in the conventional four dimensional framework. It is also expected that the higher dimensional models could naturally solve the fine tuning problem \cite{207,117,14}. 
 The effect of these higher dimensions, which are unobservable at moderate energies, could be felt at very high energies. Hence, it is natural to ask whether 
 the gravitational action is still given by Einstein-Hilbert Lagrangian. The general consensus is to modify the Einstein-Hilbert Lagrangian by the inclusion of higher curvature terms. 
 The motivation for studying models with more than three spatial dimensions comes from two key factors. First, there exists a natural class of gravitational theories in higher dimensions, which share several properties of Einstein theory in four dimensions called the Lanczos-Lovelock models which can teach us about the structure of gravitational theories. Second, it is not unbelievable that our universe does have more than three spatial dimensions with the extra dimensions either compactified or dynamically structured in an appropriate way so as to have escaped experimental detection till now \cite{232}. Meanwhile, higher dimensional brane world models have also attracted much
 attention \cite{Sheykhi3,Sheykhi1,Sheykhi2}. The complexity and richness of the structure of gravity make us more curious and motivate us to think
 deeper.
 \subsection{Gravity and thermodynamics}
 The intriguing connection between gravity and thermodynamics has emerged as an exciting research area in the past few decades. Such a connection was first demonstrated by Bekenstein and Hawking in the context of black hole mechanics. Following the semi-classical approach, it was discovered by Bekenstein and Hawking \cite{Hawking1} that black holes behave similar to a black body, emitting a characteristic wavelength known as Hawking radiation with a temperature proportional to the surface gravity at its horizon. Evidently, black holes also possess an entropy 
 proportional to the surface area of the horizon\cite{Bekenstein1}. Trailing along this line of work, the four laws of black hole thermodynamics were put forth by Carter, Hawking, and Bardeen, which turned out to be analogous to the laws of thermodynamics satisfied by ordinary macroscopic systems\cite{bardeen}. 
 This lead to the colossal recognition that black holes behave like ordinary thermodynamic objects.    
 
 The thermodynamical analysis of the black hole horizon was later extended to the cosmological horizon, for instance,
 in the case of de Sitter universe \cite{Gibbons1}, where the horizon was associated with a temperature called Hawking temperature
 $T = \frac{1}{2\pi l}$  and an entropy $S = \frac{A}{4G},$ where $ A = 4\pi l^{2}$ is the area of the cosmological event horizon with $l$ being its 
 radius. Further major step along this line was taken by Jacobson\cite{Jacob1}, who derived the Einstein's gravity equation by considering the Clausius relation, $\delta Q = TdS $  at the horizon together with the equivalence principle, $\delta Q$ and $T$ refers to the energy flux and Unruh temperature as perceived by an accelerated observer within the horizon. Possible schemes for relating gravity and thermodynamics have also been discovered for a variety of gravity theories. As shown in \cite{paddy}, the Clausius relation also comes into play while interpreting gravitational field equations as an entropy balance law, $\delta S_{m} = \delta S_{grav}$ across a null surface. These ideas also have implications in the cosmological context.  We can see that the entropy balance condition correctly reproduces the field equation - but with an arbitrary cosmological constant - without affecting the entropy balance \cite{Paddybook}. The thermodynamic perspective of the gravitational field equations has also been investigated by the authors in \cite{paddystatic,sumanta1,sumanta2,paddy5,185,cai4}.
  
 \subsection{ An emergent perspective for gravity and spacetime}

 Centuries back, scientists could explain the behaviour of matter using thermodynamic principles before understanding the molecular structure of matter. Physicists could express the behaviour of ordinary matter through some equations, which could be written in terms of macroscopic variables like density, velocity, shape, etc., without knowing the microscopic notions like the existence of molecules or atoms. However, such a thermodynamic description will involve constants like specific heat, Young's modulus, etc., which can not be calculated without knowing the underlying microscopic theory. In that case, one can work with the thermodynamic potentials like free energy, entropy, enthalpy, etc., the extremization of which could lead to the equilibrium properties of such systems. Like that, it is expected that the Emergent paradigm could describe gravity in a thermodynamic language even though the exact microscopic description of spacetime is not known to us.

 The notion of an emergent phenomenon could be explained by considering the example of the ideal gas, which is
 kept in a container with volume V. One can describe such a system through the thermodynamic relation $(P/T)\propto(1/V),$ 
 where $T$ is the temperature and $P$ is the pressure exerted by the gas on the walls of the container. It is
 possible to obtain this relation by extremizing an appropriately defined entropy functional $S(E, V )$ or the free energy,
 $F (T, V ).$ However, within the framework of thermodynamics, one can not find the reason behind the validity of such a
 relation. As emphasized by Boltzmann, the notion of temperature points at the existence of some microscopic degrees
 of freedom that can store and transfer energy in the form of heat. The continuum description of fluid can not have
 a notion of temperature since it has no internal degrees of freedom. This can be clearly understood when one obtains the
 relation, $Nk_B= PV/T,$ using the laws describing the microscopic degrees of freedom. The proportionality constant `$N$', 
 in the above relation, represents the number of microscopic degrees of freedom, which loses its significance in the
 continuum limit. Here one could count the microscopic degrees of freedom as the Avagadro number for one mole of
 gas from the thermodynamical behaviour of normal matter.

 In a similar way, one can proceed to the description of spacetime. In the past decades, it has been shown that
 the notions of temperature and entropy could be associated with the null surfaces like the black hole horizon\cite{26,Hawking1}. This
 result, which could be thought of as a unique feature of the black hole, was further generalized by Davis\cite{70}, Unruh\cite{233}, and many
 others. It has been shown that any observer in spacetime will attribute a temperature,
 \begin{equation}\label{eqn1}
 k_B T=\frac{\hbar}{c}\frac{K}{2\pi} 
 \end{equation}
 to the null surface, which he perceives as a horizon, where $K$ is the appropriately defined acceleration of the observer.
 One can consider the flat spacetime as the simplest context, where this result appears. In a flat spacetime, any
 observer having an acceleration $K,$ can feel that the spacetime is endowed with a temperature, $T=\frac{\hbar}{k_B c}\frac{K}{2\pi}.$
 This result, which was obtained for an observer moving with uniform acceleration, can also be generalized to an observer
 with a slowly varying acceleration, with $(\dot K /K^2)\ll 1.$ Since this result is more general, it is possible to introduce
 the notion of the ’Local Rindler Observers’, who perceive the spacetime as hot. Equation (\ref{eqn1}) is one of the fascinating
 relations that emerged from the principles of quantum theory and relativity. One of the major consequences
 of this result is the observer dependence of all the thermodynamic notions when one considers the non-inertial
 observer. An inertial observer attributes zero temperature to a flat spacetime, while an accelerated observer will feel
 it has a non zero temperature. Since the spacetime could be perceived by a class of observers as hot with a non
 zero temperature, one can describe the spacetime dynamics through the laws of thermodynamics as the
 dynamics of hot gas. This emergent approach to gravity and spacetime also has significant implications in the field
 of cosmology\cite{Paddy11}     
 
 \subsection{The emergent gravity paradigm-main claims}
 Some of the fascinating features of gravitational theories motivate us to interpret the gravitational field equations as
 emergent, like elasticity or fluid dynamics equations. The important claims of
 the ’emergent paradigm’ are summarized below.
 \begin{itemize}
 	\item[1] The field equations of gravity can reduce to thermodynamic identities in a variety of gravity theories which are
 	more general than Einstein gravity\cite{176,Cai_first_law_friedmann_equations,201,4,akbr_cai,179,134}.
 	\item[2] It is possible to obtain the gravitational equations of motion from the thermodynamic extremum principles \cite{195,180}.
 	\item[3] The density of microscopic degrees of freedom can be obtained through equipartition arguments\cite{paddy5}.
 	\item[4] The action functional for gravitation can have thermodynamic interpretation in a class of theories\cite{176,178,168,130}.
 	\item[5] Einstein field equation can reduce to Navier-Stokes equations when projected on any null surface in any
 	spacetime\cite{69,230,186,131}.
 	\item[6] The Euclidean path integral associated with gravitational action, interpreted as a partition function, can provide
 	expressions for free energy, energy, and entropy in Lanczos-Lovelock models\cite{129,176,Gibbons1}.
 	\item[7] It is possible to interpret the gravitational action as a momentum space action in terms of a pair of conjugate
 	variables, the variation of which has a natural thermodynamic interpretation\cite{203}.
 	\item[8] The emergent paradigm provides a solution to the most profound puzzle, the cosmological constant problem\cite{193,194}. 
 	\item[9]
 	It is possible to interpret the total Noether charge inside a volume $R$, linked to the time evolution vector field, as the heat content
 	of the boundary $ \partial R$ of the volume. It is also possible to describe the evolution of spacetime in a purely thermodynamic language in terms of the appropriately defined degrees of freedom. The time evolution of the spacetime geometry can be interpreted as the heating and cooling of  null surfaces, which is driven by the departure from the holographic equipartition \cite{190}.
	\item[10]  The quantum corrected field equations of gravity has been obtained from 
		the direct link between the effective action for gravity and the kinetic theory of the spacetime fluid\cite{paddymesoscop}.	
	
	 \end{itemize}
 
 \subsection{Evolution of the universe as the emergence of cosmic space}
 
The basis of the emergence of cosmic space is the holographic principle, which suggests a relation between the number of degrees of freedom within a given volume of space with that residing on the boundary surface. Cosmological observations indicate that our universe is proceeding to a final de Sitter state \cite{I28,I39,I37,I35,Zhao1}. 
A pure de Sitter universe obeys 
the holographic principle in the form,
 \begin{equation}\label{equipart}
 N_{surf} = N_{bulk},
 \end{equation}
 where $N_{surf}$ is the number of degrees of freedom on the surface of the Hubble sphere 
 and  $N_{bulk}$ is the number of degrees of freedom residing in the bulk region of space
 enclosed by the horizon. 
 The surface degrees of freedom can be defined as,
 \begin{equation}\label{eqn:Nsurf1}
 N_{surf} = \frac{4 \pi}{ L_{P}^{2} H^2}.
 \end{equation}
 Here, we have chosen $'L_{P}^2=\frac{G\hbar}{c^3}'$ to denote one degree of freedom since $L_{P}$ act as the lower bound of length scales that can be operationally defined\cite{Paddy2}. 
 
 The bulk degrees of freedom is given by,
 \begin{equation}\label{eqn:Nbulk1}
 N_{bulk} =\frac{\mid E \mid}{\frac{1}{2} {k_B T }},
 \end{equation}
 where $E=(\rho+3p)V$, the Komar energy inside the Hubble volume, $V=\frac{4\pi}{ 3 H^3}$; $k_B$, the Boltzmann constant and $T = \frac{H}{2\pi}$ is
 the Gibbon's Hawking temperature. In other words, $N_{bulk}$ can be defined as the number of microscopic degrees of freedom required to store the bulk gravitating energy $(\rho+3p)V $ at temperature $T=H/2\pi$ which is given by the equipartition law $N= \frac{E}{1/2 k_{B}T}$. It is also possible to obtain equations which are similar to law of emergence, with other definitions of temperature, which contain temporal derivatives of Hubble parameter. But, the resulting equations does not seem to have any simple interpretation \cite{Paddy1}. It is to be noted that $N_{bulk},$ a dimensionless number associated with the bulk energy $E_{Komar},$ is the bulk degrees of freedom
 which are in equipartition at the temperature $T$. It is important that, the temperature $T$  
 should not be confused with the 
 normal kinetic temperature of the matter. Similarly, the degrees of freedom $N_{bulk}$ are not the standard degrees of freedom of matter, instead, 
 they  
 are those which have already emerged from some pre geometric variables, along with the space. 
 Analogically it is possible to treat the degrees of freedom $N_{bulk}$ that have emerged 
 in the bulk, as though they are inside a microwave oven having a temperature equal to 
 the surface.

 From the equations (\ref{equipart}) and 
 (\ref{eqn:Nbulk1}), we can have 
 $\mid E \mid={{1\over2}{k_B T N_{surf}}}$. This relation  
 is known as the holographic equipartition since it relates the degrees of freedom in the bulk 
 determined by the equipartition condition to the degrees of freedom on the horizon surface. Observations indicate that our universe is evolving
 towards a pure de Sitter phase that satisfies holographic equipartition. The
 evolution of the universe can then be considered equivalent to the emergence
 of space due to its tendency to achieve holographic equipartition. In other words, it is 
 the holographic discrepancy, `$N_{surf} - N_{bulk},$' which drive the accelerated expansion of the universe. The most natural form of such a law will be,
 \begin{equation}
 \Delta V={\Delta t(N_{surf}-\epsilon N_{bulk})} ,
 \end{equation}
 where $V$ is the Hubble volume and $t$ is the cosmic time, both in Planck units. This equation can be reframed suitably, by setting $\Delta V/\Delta t= dV/dt$ and reintroducing Planck length, to a form,
 \begin{equation} \label{eqn:dVdt1}
 {dV\over dt} ={L_{P}^{2}(N_{surf}- \epsilon N_{bulk})},
 \end{equation}
  where $\epsilon = +1$ if $(\rho+3p)<0$ and $\epsilon=-1$ if $(\rho+3p)>0.$ This relation can be considered as a postulate governing the emergence of space. Padmanabhan have  
 shown that, the above expression reduces to the Friedman equation. Thus, using the concept of holographic equipartition, the evolution of the universe can be described as the emergence of cosmic space with the progress of cosmic time.  
 
 Cai \cite{Cai1} extended Padmanabhan's proposal to (n+1) dimensional FRW universe in the context of Einstein's gravity and more general gravity theories like Gauss-Bonnet and Lovelock models. However, Cai's work was criticised for using the so-called effective volume to find the rate of change of volume, while the simple ordinary volume is used for defining the bulk degrees of freedom. To overcome this discrepancy, Yang et. al.\cite{Yang} have used the plain volume to determine both the rate of change of volume and the bulk degrees of freedom, and reformulated the law of emergence accordingly. Later Sheykhi \cite{sheykhi} extended the law of emergence to  the non-flat universe in (n+1) Einstein, Gauss-Bonnet, and Lovelock gravity theories. This idea of emergence has also been extended to the braneworld scenarios \cite{Sheykhi1,Sheykhi2,Sheykhi3}. For more investigations on this idea, see 
 \cite{sumanta1,sumanta22,Komatsu,Zhang,Hashemi,MKT,HAKT,HAKT2,KT3,Komatsu1}.

 \section{Law of emergence from first law of thermodynamics}
 
The exciting feature of the law of emergence is that it  reduces to the Friedmann equation in Einstein's and more general gravity theories.
For example, in the context of (3+1) Einstein's theory, knowing the volume of the horizon as, $V=4\pi/3H^3,$ the left hand side of the law of emergence can be found to be, $dV/dt \propto (-\dot H/H^4).$ Now, substituting for $N_{surf} \propto H^{-2}$ and 
$N_{bulk}= 2 (\rho+3p) V/T,$ and using the relation for temperature, $T=H/2\pi,$ the law of emergence can easily be reduced to,
 \begin{equation}
 \frac{\ddot a}{a}=\frac{4\pi L_{P}^2}{3} \left(\rho+3p\right),
 \end{equation}
 which is the space-space component of Einstein's equation describing the dynamics of the FRW universe. From this, with the help of the conservation equation, $\rho da^3 = - p da^3,$ it is possible to obtain the first Friedmann equation, $H^2=(8\pi G/3) \rho.$ In the conventional approach, Friedmann equations arise as the special solution
to the field equation of gravity. Compared to this the approach of the law of emergence in describing the evolution of the universe is appeared to be simple. Due to its overwhelming simplicity and the conceptual difference in explaining
the evolution of the universe, it is advantageous to motivate the law of emergence
from some well established fundamental principles, rather than taking this law as granted. In this section, we aim to establish the connection of the law of emergence proposed by Padmanabhan and its different generalisations with the first law of thermodynamics. 
We use the unified first law \cite{sheykhi,Cai1,Yang}, $dE=TdS+WdV,$ which has the same form regardless of the gravity theories, to deduce the law of emergence, 
which takes different forms in different gravity theories.  
One may suspect the novelty of this derivation, since the Friedmann equations
can be cast as the first law of thermodynamics \cite{caiakb}, at the apparent horizon. But, this does not be an ambient reason to conclude that 
the law of emergence is  
equivalent to the first law of thermodynamics. Because,  
not all cosmological models that satisfy the Friedmann equation are consistent with the law of emergence, since it
imposes an additional constraint,
$N_{surf} \geq N_{bulk}.$ 
In addition, our aim is to deduce the law of emergence from the first law of thermodynamic is similar in spirit of the work of Jacobson \cite{Jacob1}, who derived the Einstein's field equation from the thermodynamical principle. 
Below we arrive at all these forms of the law of emergence by starting from the fundamental thermodynamic principle, the first law of thermodynamics.

\subsection{Padmanabhan's law of emergence from first law of thermodynamics}

Here we are deducing the law of emergence due to Padmanabhan from the first law of thermodynamics. For an expanding universe, it is viable to attribute thermodynamic properties like temperature and entropy, 
to the event horizon of a de Sitter universe \cite{GH}. If  matter with energy $dE$ passes through such a horizon, then we have $-dE=TdS.$ But in non-de Sitter or even quasi de Sitter spacetimes, an event horizon does not exist, 
instead there exists an apparent horizon. In reference \cite{Frolov2003} the authors have used the above thermodynamic principle, i.e. $-dE=T dS$ to calculate the energy flux through such an apparent horizon and derived the 
Friedmann equations. 
Meantime, Bousso has argued that a thermodynamic description is approximately valid for the apparent horizon too \cite{Bousso2005}  and have shown that the first law of thermodynamics is indeed hold at the apparent horizon. Moreover, in reference \cite{Bak2000,Hayward1998, Hayward1999}, the authors have shown that, for a dynamical spacetime, the apparent horizon can be attributed with the gravitational entropy and surface gravity, a measure of temperature, like a causal horizon. Hence like an event horizon, thermodynamic identities are valid for apparent horizons too. 

 The Friedmann equations of FRW universe are the field equations with a source of perfect fluid, for which there exists a well defined pressure $p$ and density $\rho.$ Hence it is more natural to use the thermodynamic principle of the form, $TdS=dE+PdV$ to establish the thermodynamic connection of gravity. Motivated by this, it was shown that, in cosmological set up one can use a first law of thermodynamics of the form, $dE=TdS+WdV,$ where $W=(\rho-p)/2,$ the work density, which on projecting to the apparent horizon of the expanding universe will naturally reduces to Friedmann equation. In this equation the energy $E=\rho V$ is the Misner-Sharp energy enclosed by the horizon, $S$ is the horizon entropy proportional to the apparent horizon area and $T=\kappa/2\pi$ is the horizon temperature with $\kappa$ as the surface gravity of the apparent horizon. 
 
 While formulating the unified first law, $dE = TdS +WdV$ , the energy $E$ is taken as the Misner-Sharp energy enclosed by the horizon, which is at temperature, $T=\kappa/2\pi$. But, while defining the bulk degrees of freedom we have to consider the bulk gravitating energy, $(\rho +3p)V$. Padmanabhan defined $N_{bulk}$ as the number of  microscopic degrees of freedom required to store the bulk gravitating energy $(\rho +3p)V$ at temperature $T= H/2\pi$, which is given by the equipartition law $N= \frac{E}{1/2 k_{B}T}$. It has to be noted that $'E'$ and $ 'T' $ in the unified first law and the law of emergence are not the same, although we have used the same notation in both cases (as per convention). In the cosmological context, the temperature $T$ in a thermodynamic relation is generally defined in accordance with the definition of energy $E$, while the entropy, $S$ depends only on the theory of gravity. For example, see the first law in the form $-dE = TdS$, where $-dE= A(\rho +p)r_{A}Hdt$ is the energy flux through the apparent horizon, which is kept at temperature $T=1/2\pi r_{A}$. It is also possible to reach at the law of emergence and its generalized versions from the first law of the form, $-dE = TdS$, where $T=1/2\pi r_{A}$. For details see \cite{HKT}. It may be possible to define the bulk degrees of freedom, which are in equipartition at horizon temperature $T=\kappa/2\pi$. However, if we then choose $E=(\rho +3p)V$, where $V$ is the Hubble volume, the resulting equation of emergence will not have a simple interpretation. If we choose $T=H/2\pi$ and $V= 4\pi/3H^{3}$ to define the bulk degrees of freedom, required to store the gravitating energy $(\rho +3p)V$, the expansion of the universe could be interpreted as the emergence of cosmic space through a very simple equation $ {dV\over dt} ={L_{P}^{2}(N_{surf}- \epsilon N_{bulk})},$ which naturally generates the Friedmann equation. It is why Padmanabhan had chosen $T=H/2\pi$. For more details see \cite{Paddy1}. 
 

Here we present the derivation of the law of emergence from the thermodynamic principle for a flat universe. We will first look at the derivation of the law proposed by Padmanabhan, which is for a flat, (3+1) dimensional Einstein's gravity. 
We will use the 
the unified first law of thermodynamics given by\cite{caiakb},
\begin{equation}\label{eqn:tdsunified}
dE=TdS+WdV .
\end{equation}
The variation in energy can be written as,
\begin{equation}
dE=-\left(\frac{4\pi}{H^2}\right)(\rho+p) dt - \left(\frac{4\pi}{H^4}\right) \rho \dot H dt  ,
\end{equation}
where we have used the conservation equation, $\dot\rho+3H(\rho+p)=0.$ The entropy of the horizon is given by the Bekenstein relation \cite{Bekenstein1},
\begin{equation}\label{entropy1}
	S=\frac{A}{4L_{P}^2},
\end{equation}
where $A=4\pi/H^2,$ the horizon area of flat universe and $L_{P}$ is the Planck length. The temperature of the horizon can be expressed as\cite{Cai_first_law_friedmann_equations},
\begin{equation}\label{Teqn}
T=\frac{-H}{2\pi} \left(1+\frac{\dot H}{2H^2}\right).
\end{equation}
It is to be noted that this temperature vanishes for the radiation dominated universe since the surface gravity vanishes in the radiation dominated epoch with equation of state parameter $\omega= 1/3 $ \cite{Tian}.	
In that case, the heat supply term, $ TdS $ in the unified first law  vanishes and the variation in the total energy $dE$ will be equal to the work term, $WdV$.

 Now, using the above equations, the unified first law in equation (\ref{eqn:tdsunified}) can be expressed as,
\begin{equation}
\begin{split}
-\left(\frac{4\pi}{H^2}\right)(\rho+p) dt - \left(\frac{4\pi}{H^4}\right) \rho \dot H dt = \frac{-H}{2\pi} \left(1+\frac{\dot H}{2H^2}\right) \left( \frac{1}{4L_{P}^2}\right) \left(-\frac{8\pi}{H^3} dH\right) + \\
\left(\frac{\rho-p}{2}\right) \left(\frac{-4\pi}{H^4}\right) dH. 
\end{split}
\end{equation}
The above equation can be simplified to, 
\begin{equation}
-\frac{4\pi}{H^4} \dot H = \left(\frac{4\pi}{H^4}\right) 4\pi L_{P}^2 (\rho+p), 
\end{equation}
which can further be expressed as,
\begin{equation}
-\frac{4\pi}{H^4} \dot H=\left(\frac{4\pi}{H^4}\right) L_{P}^2 \left(\frac{8\pi}{3}\rho + \frac{8\pi}{6} (\rho+3p) \right).
\end{equation}
Substituting for $\rho$ in the first term inside the second parenthesis on R.H.S. and using Friedmann equation, $H^2 = (8\pi/L_{P}^2)\rho$ we arrive at, 
\begin{equation}
-\frac{4\pi}{H^4} \dot H = L_{P}^2 \left( \frac{1}{L_{P}^2} \frac{4\pi}{H^2} + \frac{(\rho+3p) V}{\frac{1}{2} (H/2\pi)}\right).
\end{equation}
It is easy to identify the left hand side as the rate of change of the horizon volume, consequently the above equation reduces to the law of emergence as proposed by Padmanabhan, 
\begin{equation}
\frac{dV}{dt} = L_{P}^2 \left(N_{surf} - N_{bulk} \right),
\end{equation}
where the degrees of freedoms can be identified as,
\begin{equation}\label{dofms1}
N_{surf}=\frac{A}{L_{P}^2} \quad N_{bulk}=-\frac{(\rho+3p) V}{\frac{1}{2} (H/2\pi)}.
\end{equation}
The degrees of freedom should always be positive, hence it appears that the above definition of $N_{bulk}$ makes sense only for accelerating epoch of the universe at which $(\rho+3p)<0.$ But in the prior decelerated epoch, in which the normal matter would be the dominant component, the term $(\rho+3p)>0$ there by makes $N_{bulk}<0.$ This discrepancy can be alleviated by using appropriate sign by writing the law of emergence as given in equation (\ref{eqn:dVdt1}).
Here the bulk degrees of freedom is,
\begin{equation}\label{equipart1}
 N_{bulk}=-\epsilon \frac{(\rho+3p) V}{\frac{1}{2} (H/2\pi)}.
\end{equation}
The procedure described above have shown that, like the deduction of Einstein's field equations from the thermodynamic identity, it is possible to obtain the law of emergence proposed by Padmanabhan from the first law of thermodynamics. 

\subsection{Cai's generalisation of law of emergence from first law of thermodynamics}
The original proposal of the law of emergence by Padmanabhan, was extended to (n+1) dimensional flat universe with $n>3,$ by Cai \cite{Cai1}. In this case, the degrees of freedoms are defined as \cite{verlinde},
\begin{equation}\label{dofn}
	N_{surf}=\frac{\alpha A}{L_{P}^{n-1}}  \quad N_{bulk}=\frac{-4\pi\Omega_n}{H^{n+1}}\frac{(n-2)\rho+np}{n-2},
\end{equation}
where $\alpha=(n-1)/2(n-2)$ and  $A=n\Omega_n/H^{n-1}$ is the area of the $n-$dimensional space, with $\Omega_n$ as the volume of the unit $n-$sphere. Here $N_{bulk}$ is defined using the original relation $N_{bulk}= -2E_{Komar}/T,$ but with Komar energy $E_{Komar}= [((n-2)\rho+np)/(n-2)]V$ for the (n+1) dimensional universe. Following these, the law of emergence for a flat universe of (n+1) dimension is proposed as \cite{Cai1},
\begin{equation}\label{emergndimen}
\alpha \frac{dV}{dt} = L_{P}^{n-1} \left(N_{surf} - N_{bulk} \right),
\end{equation}
where $V=\Omega_n/H^n,$ is the volume of the $n$ sphere. Using this form of the law, Cai has derived the Friedmann equation,
\begin{equation}\label{friedn+1}
H^2 = \frac{16\pi L_{P}^{n-1} }{n(n-1)} \rho, 
\end{equation}
of flat (n+1) dimensional FRW universe.

Following the procedure in the previous subsection, for 
deducing the law of emergence given in equation (\ref{emergndimen}) from the first law, we need the variations of the actual energy, $E$ within the Hubble horizon, entropy of the horizon and of course that of the volume of the horizon. These quantities for (n+1) dimensional flat universe are defined as, $E=\rho V$, entropy of the horizon is \cite{Bekenstein1,wald}
\begin{equation}\label{entrond}
S= n\Omega_n/4L_{P}^{n-1}H^{n-1},
\end{equation}
 and relation for volume is as given in the previous paragraph. On feeding these variations in to the unified first law and using the continuity equation, $\dot\rho+nH(\rho+p)=0$ we can arrive at,\cite{MKT} 
\begin{equation}\label{reduced_firstlaw}
-\frac{n\Omega_n \dot H}{H^{n+1}} = \frac{n\Omega_n}{H^{n+1}} \frac{8\pi L_{P}^{n-1} \left(\rho+p\right)}{n-1}.
\end{equation}
The left hand side of the above equation can be identified as $dV/dt.$ Splitting  the term on the right hand side as 
$n(\rho+p)=\left[(n-2)\rho+np\right]+2\rho$ then by some suitable rearrangements, using
Friedmann equation in (\ref{friedn+1}) we finally arrive exactly at the law of emergence given in equation (\ref{emergndimen}). 
In reaching this result, the degrees of freedom can be suitably defined as given in equation(\ref{dofn}).

\label{sec:chp7_3}
Let us next consider the case of 
Gauss-Bonnet gravity. The crucial point in Cai's proposal is in the definition of the surface degrees of freedom.  Considering the entropy formula in Gauss-Bonnet gravity, which no longer obeys the Bekenstein-Hawking area formula \cite{Cai2,Cai3}, 
\begin{equation}\label{7equ:S}
S = \dfrac{A}{4L_{P}^{n-1}}\Bigg(1+\dfrac{n-1}{n-3}\dfrac{2\tilde{\alpha}}{H^{-2}}\Bigg),
\end{equation}
where $\tilde{\alpha}=(n-2)(n-3)\alpha$, the Gauss-Bonnet coefficient, an effective area can be defined as, 
\begin{equation}\label{7eqn:area}
\tilde{A} =A\Bigg(1+\dfrac{n-1}{n-3}\dfrac{2\tilde{\alpha}}{H^{-2}}\Bigg),
\end{equation}
where $A=n\Omega_n/H^{n-1}.$ There exists a volume $\tilde{V}$ corresponding to this. Using the standard relation between the volume and area of $n$-sphere, Cai has arrived at the rate of increase of the volume $\tilde{V}$ corresponding to the area, $\tilde{A}$ as,  
\begin{equation}
\frac{d\tilde{V}}{dt} = -\frac{n\Omega_n}{H^{n+1}}\left(1+2\tilde{\alpha}H^2 \right) \dot H.
\end{equation}
Making use of this relation a choice of surface degrees freedom has been made as, 
\begin{equation}\label{nsurfgb}
N_{surf} =\dfrac{\alpha n\Omega_n}{H^{(n+1)}L_{P}^{n-1}} \left(H^2 +\tilde{\alpha} H^4 \right).
\end{equation}
Nevertheless, the bulk degrees of freedom is still defined using relation, $N_{bulk}=-2E_{Komar}/T,$ where the ordinary volume $V$ has been used and similarly for temperature we still follow the standard relation. This choice of different relations for volumes in the case of surface and bulk degrees of freedoms has been criticised \cite{Yang}, about which we will discuss in later section and for time being we will follow Cai's approach.

 Now, for deducing the law of emergence from the first law, we have to substitute the variation of entropy given equation (\ref{7equ:S}), variation of $E=\rho V$ and  unlike entropy, 
the work density, $W=(\rho-P)/2$ into the unified first law. The resulting equation, with the help of the corresponding Friedmann equation \cite{Cai_first_law_friedmann_equations,citekey},
\begin{equation}\label{eqn:sheykhi4}
H^{2}+\tilde{\alpha}H^{4} = \frac{16\pi L_{P}^{n-1}}{n(n-1)}\rho
\end{equation}
can be reduced to the required form,
\begin{equation}\label{7eqn:sh}
\alpha \dfrac{d\tilde{V}}{dt} = L_{P}^{n-1} \Big(N_{surf} - N_{bulk}\Big),
\end{equation}
which is the 
expansion law 
proposed by 
Cai in order to derive Friedmann equation of a flat FRW universe in Gauss-Bonnet gravity.

Now we consider the 
Lovelock gravity theory which generalizes Einstein gravity when spacetime have dimension greater than four \cite{Love1,Lanczos}. The entropy associated with the apparent horizon in this case can be defined as \cite{cai_love},
\begin{equation}\label{7entropy}
S=\frac{A}{4L_{P}^{n-1}}\sum_{i=1}^{m} \frac{i(n-1)}{n-2i+1}\hat{c_{i}}H^{-(2-2i)},
\end{equation}
where $m=[n/2]$ and the coefficients $\hat{c_{i}}$ are given by,
\begin{equation}
\begin{aligned}\label{7eqn:ci}
\hat{c_{0}}=\frac{c_{0}}{n(n-1)}, \,   \hat{c_{1}}=1,  \, \, 
\hat{c_{i}}=c_{i} \prod_{j=3}^{2m}(n+1-j)\hspace{2mm}   when  \hspace{5mm} i>1.
\end{aligned}
\end{equation}
Defining an effective area $\tilde{A} =4L_{P}^{n-1}S$, 
and a rate of increase of the effective volume 
\cite{Cai1} as $d\tilde{V}/dt={(1/H^{-1}(n-1))} {d\tilde{A}/dt}$,
an effective surface degrees of freedom can be defined as
\begin{equation}\label{lovedofflat}
N_{surf}=\frac{\alpha n\Omega_{n}H^{-(n+1)}}{L_{P}^{n-1}} \sum_{i=1}^{m}c_{i}H^{2i}.
\end{equation}  
By using the standard definition of the bulk degrees of freedom,  
Cai proposed a 
modified expansion law,
\begin{equation}\label{7eqn:lovelockexp}
\alpha \dfrac{d\tilde{V}}{dt} = L_{P}^{n-1} (N_{surf} - N_{bulk}).
\end{equation}
One should notice that, even though, the above form of the law of emergence is similar to that for the Gauss-Bonnet in the present case, the entropy used to extract the degrees of freedom is completely different. So the resemblance in form is only superficial.

Below we 
show that equation (\ref{7eqn:lovelockexp}) also can be derived from the first law (\ref{eqn:tdsunified}).
Assuming the standard 
form of $E, W$ and $T,$ 
same 
as used previously 
and entropy 
of the form given in (\ref{7entropy}), one can rewrite the first law (\ref{eqn:tdsunified}) as,
\begin{equation}\label{7eqn:llg11}
n\Omega_{n}H^{-(n+1)} (-\dot H/H^2) \sum_{i=1}^{m}ic_{i}H^{2i}=L_{P}^{n-1}\Bigg(\frac{N_{surf}}{\alpha}-\frac{N_{bulk}}{\alpha}\Bigg).
\end{equation}
From the entropy relation (\ref{7entropy}), the effective area of the apparent horizon is given by \cite{sheykhi},
\begin{equation}\label{effectiveA}
\tilde{A}=n\Omega_{n}H^{1-n}\sum_{i=1}^{m} \frac{i(n-1)}{n-2i+1}\hat{c_{i}}H^{-(2-2i)} ,
\end{equation}
and thus 
the effective volume of the apparent horizon in Lovelock gravity can be expressed as
\begin{equation}
\frac{d\tilde{V}}{dt}=n\Omega_{n}H^{-(n+1)} (-\dot H/H^2) \sum_{i=1}^{m}ic_{i}H^{2i}.
\end{equation} 
Now, it is straightforward to identify 
L.H.S. of equation (\ref{7eqn:llg11}) as the rate of change of the effective volume.
As a result, equation (\ref{7eqn:llg11}) becomes,
\begin{equation}\label{7eqn:lovelock}
\alpha\frac{d\tilde{V}}{dt}=L_{P}^{n-1}\Big(N_{surf}-N_{bulk}\Big).
\end{equation}
This is the law of emergence proposed by Cai \cite{Cai1} for deriving the Friedmann equation for a flat FRW universe in Lovelock gravity.

\subsection{Yang et. al.'s version of the law of emergence from first law of thermodynamics}

 In Cai's generalization of the law of emergence, the number of degrees of freedom on the holographic surface $N_{sur}$ and the rate of change of the spatial volume
 were modified, by taking account of the entropy formulas in the respective gravity theories.  
Yang et al \cite{Yang} have proposed another generalization of the Padmanabhan's original version of the law of emergence, for flat universe. In this approach the $N_{surf}$ is taken to be proportional to the area of the horizon irrespective of the gravity theory, and $N_{bulk}$ is obtained using the equipartition law of energy. Following this proposal, the emergence of space is described by the dynamical equation of the form,
\begin{equation}\label{yangeq}
\frac{dV}{dt} = L_{P}^{n-1} f (\Delta N, N_{surf}), 
\end{equation} 
where $\Delta N = N_{surf} - N_{bulk}.$ Yang et. al. have assumed $f(\Delta N, N_{surf})=\Delta N / \alpha$ for flat universe in Einstein's gravity. While for Gauss-Bonnet gravity it is assumed as,
\begin{equation}\label{eqn:yang1}
f(\Delta N, N_{surf}) = \frac{\Delta N / \alpha +\bar{\alpha} K (N_{surf} / \alpha)^{1+\frac{2}{1-n}}}{1+2\bar{\alpha} K (N_{surf}/\alpha)^{2/1-n} },
\end{equation}
where $K=(n\Omega_n/L_{P}^{n-1})^{\frac{2}{n-1}} $, and $\bar{\alpha}$ is a parameter with length dimension 2. In the case of Lovelock gravity, the above function is assumed to be of the form,
\begin{equation}\label{fdeltaN}
f(\Delta N, N_{surf}) = \frac{ \Delta N / \alpha + \sum_{i=2}^m \bar{c}_i K_i (N_{surf}/ \alpha)^{1+\frac{2i-2}{1-n}} } {1+\sum_{i=2}^m i \bar{c}_i K_i (N_{surf}/\alpha)^{\frac{2i-2}{1-n}} },
\end{equation}
where $K_i=K=(n\Omega_n/L_{P}^{n-1})^{\frac{2i-2}{n-1}}, m=n/2 $ and $\bar{c}_i$ are coefficients having dimension $(2i-2).$ 

We will now show that Yang et. al.'s version of the law of emergence, can also be derived from the more fundamental thermodynamic identity given in equation (\ref{eqn:tdsunified}). 
First consider the case of Gauss-Bonnet gravity. 
Let us consider 
(\ref{reduced_firstlaw}) (for the flat case), which we have obtained from the unified first law for Gauss-Bonnet gravity. This equation can be written in slightly different form, after substituting the the respective Friedmann equation as,
\begin{equation}\label{7eqn:tds4}
n\Omega_{n}H^{-(n+1)}\Big[-\dot{H}(1+2\tilde{\alpha}H^{2})\Big] =n\Omega_{n}H^{-(n+1)}\Big[ H^{2} + \dfrac{8\pi  L_{P}^{n-1}((n-2)\rho + np)}{n(n-1)} + \tilde{\alpha}H^{4}\Big].
\end{equation}
This expression can be suitably re-arranged in such a way that, the term 
$dV/dt=-n\Omega_{n}\dot{H}H^{-(n+1)},$ corresponding the change in volume, and then  
identifying the degrees of freedoms as, 
\begin{equation}\label{7eqn:nsur12}
N_{sur} = \dfrac{\alpha\hspace{1mm} n\Omega_{n}H^{-(n-1)}}{L_{P}^{n-1}} = \dfrac{\alpha A}{L_{P}^{n-1}}, \quad
N_{bulk}  = -\dfrac{4\pi V H^{-1}[(n-2)\rho + np]}{n-2},
\end{equation}
and a constant,
$K=(n\Omega_{n}/L_{P}^{n-1})^{2/(n-1)}. $
As a result, equation (\ref{7eqn:tds4}) can be cast in to a more compact form as,
\begin{equation}\label{7eqn:qwerty}
\dfrac{dV}{dt} = L_{P}^{n-1} \left(\frac{\Delta N / \alpha +\bar{\alpha} K (N_{surf} / \alpha)^{1+\frac{2}{1-n}}}{1+2\bar{\alpha} K (N_{surf}/\alpha)^{2/1-n} } \right).
\end{equation}
Comparing with equation (\ref{eqn:yang1}), it is evident that the above equation is indeed the 
modified expansion law due 
to  Yang et.al.\cite{Yang} for Gauss-Bonnet gravity.

We will now show that the law of emergence in \cite{Yang} for Lovelock gravity will also readily follows from the first law, $dE=TdS+WdV$, just like in the 
previous case.
Beginning with the first law and following the same steps as before, after some simplifications, 
we arrive at,
\begin{equation}
\begin{split}
\frac{dV}{dt}\Bigg[1+\sum_{i=2}^{m}i\hat{c_{i}}\Bigg(\frac{n\Omega_{n}}{L_{P}^{n-1}}\Bigg)^{\frac{2i-2}{n-1}}\Bigg(\frac{n\Omega_{n}H^{-(n-1)}}{L_{P}^{n-1}}\Bigg)^{\frac{2i-2}{1-n}}\Bigg]=L_{P}^{n-1}\Bigg[\frac{n\Omega_{n}H^{-(n-1)}}{L_{P}^{n-1}}+  \Bigg. \\ \Bigg. \frac{\Omega_{n}H^{-(n+1)}8\pi[(n-2)\rho+np]}{n-1}+ \sum_{i=2}^{m}\hat{c_{i}}\Bigg(\frac{n\Omega_{n}}{L_{P}^{n-1}}\Bigg)^{\frac{2i-2}{n-1}}\Bigg(\frac{n\Omega_{n}H^{-(n-1)}}{L_{P}^{n-1}}\Bigg)^{1+\frac{2i-2}{1-n}}\Bigg] .
\end{split}
\end{equation}
We will now identify both $N_{sur}$ and $N_{bulk}$ in the above equations as, 
\begin{equation}\label{7eqn:nsur}
N_{sur}=\frac{\alpha n\Omega_{n}H^{-(n-1)}}{L_{P}^{n-1}} ,
\end{equation}
\begin{equation}
N_{bulk}= -\dfrac{4\pi V H^{-1}[(n-2)\rho + np]}{n-2}.
\end{equation}
In the present case we define the constant $K$ as, 
$K = (n\Omega_n/L_{P}^{n-1})^{(2i-2)/(n-1)} $
and then we easily recover Yang et. al.'s proposed relation,
\begin{equation}\label{7eqn:103}
\frac{dV}{dt}=L_{P}^{n-1}\Bigg[\frac{\Delta N/\alpha+\sum_{i=2}^{m}\hat{c_{i}}k_{i}(N_{sur}/\alpha)^{1+\frac{2i-2}{1-n}}}{1+\sum_{i=2}^{m}i\hat c_{i}k_{i}{(N_{sur}/\alpha)}^{\frac{2i-2}{1-n}}}\Bigg] .
\end{equation}
All these 
confirms  
that the modified versions of Padmanabhan's original conjuncture can 
be derived from the first law of thermodynamics.

To summarise, in literature there are various forms of the law of emergence that give the same Friedmann equations in different gravity theories \cite{Cai1,Yang}. 
We have shown that all these forms of the law of emergence, irrespective of the gravity theory for 
flat 
universe, can be derived from 
the first law of thermodynamics, $dE = TdS + WdV$, a well established fundamental principle.  

\subsection{Law of emergence for non-flat universe from First law of thermodynamics}

The law of emergence was extended to non-flat universe mainly by Sheykhi \cite{sheykhi}. Following the method developed in the previous section, it is easy to show that the law of emergence in non-flat universe can be obtained from the First law of thermodynamics in the context of Einstein, Gauss-Bonnet and more general Lovelock gravity models. The important change in the case of the non-flat universe is 
the use of apparent horizon radius\cite{caiakb} given by,
\begin{equation}\label{7eqn:apparent}
r_{A}=\frac{1}{\sqrt{H^{2}+\frac{k}{a^{2}}}}.
\end{equation}  
The corresponding degrees of freedom takes the form,
\begin{equation}\label{deFms}
N_{surf} = 4S=\frac{4\pi r_A^2}{L_{P}^2}, \quad 
N_{bulk} = \frac{2 \mid E_{Komar}\mid }{T} .
\end{equation}
Having these, Sheykhi \cite{sheykhi} proposed a modified version for the law of emergence for a non-flat universe in the context of (3+1) Einstein's gravity as, 
\begin{equation}
\frac{dV}{dt} = L_{P}^2 r_A H \left(N_{surf} - N_{bulk} \right).
\end{equation}
While for (n+1) dimension, the law of emergence takes a slightly different form,
\begin{equation}\label{loen+1}
\alpha \frac{dV}{dt} = L_{P}^{n-1} r_A H \left( N_{surf}-N_{bulk}\right) .
\end{equation}

We will first obtain the law of emergence for (n+1) universe.
For an (n+1) dimensional spacetime, we need the energy, entropy and temperature of the horizon to formulate the first law. The energy within the apparent horizon is, 
\begin{equation}\label{7eqn:e1}
E = \Omega_{n}r_{A}^{n} \rho ,
\end{equation}
and the entropy of the apparent horizon is proportional to its 
surface 
area, which is given as \cite{Bekenstein1,wald},
\begin{equation}\label{entroN}
S =\frac{n\Omega_{n}r_{A}^{n-1}}{4L_{P}^{n-1}}. 
\end{equation}
Meanwhile the temperature of the horizon, which is a measure of its surface gravity, $\kappa$ can be expressed as, 
\cite{caiakb},
\begin{equation}\label{7eqn:temp}
T = \frac{\kappa}{2\pi} = \frac{1}{2\pi} \Big[\frac{-1}{r_{A}}\Big(1-\frac{\dot{r}_{A}}{2Hr_{A}}\Big)\Big].
\end{equation}
Using the above equations, 
the product $TdS$ can be obtained as,
\begin{equation}\label{eqn:TdS1}
T dS=\frac{1}{2\pi} \Bigg[\frac{-1}{r_{A}}\Big(1-\frac{\dot{r}_{A}}{2Hr_{A}}\Big)\Bigg]\frac{n(n-1)\Omega_{n}r_{A}^{n-2}dr_{A}}{4L_{P}^{n-1}}.
\end{equation}
Substituting the variations $dE$ and $dS$ into the unified first law and following the similar procedure in the previous section it is now easy to deduce the law of emergence as given in equation (\ref{loen+1}).
In the case of a flat universe ($k=0$), the apparent horizon reduces to the Hubble horizon. Then, one can reach at the expansion law, which is postulated by Cai for a flat universe in (n+1) Einstein's gravity \cite{Cai1}, following the same procedure.

The above procedure can easily be extended to Gauss-Bonnet and more general Lovelock gravity theories. In the case of Gauss-Bonnet gravity, the entropy and effective area of the horizon for the non-flat universe can be obtained from equations (\ref{7equ:S}) and (\ref{7eqn:area}), respectively, 
by just replacing $H^{-1}$ with $r_A.$ 
Using the corresponding Friedmann equation for Gauss-Bonnet gravity \cite{Cai_first_law_friedmann_equations}, 
\begin{equation}\label{eqn:sheykhi4}
H^{2}+\frac{k}{a^{2}}+\tilde{\alpha}\Bigg(H^{2}+\frac{k}{a^{2}}\Bigg)^{2} = \frac{16\pi L_{P}^{n-1}}{n(n-1)}\rho
\end{equation}
 the first law can be properly reduced to the law of emergence after a little manipulation. The expressions of the degrees of freedom in this case can be realised from equation (\ref{nsurfgb}) by replacing $H$ with $r_A^{-1}.$

For Lovelock gravity, 
the law of emergence in non-flat universe takes the form, 
\begin{equation}\label{7eqn:sh}
\alpha \dfrac{d\tilde{V}}{dt} = L_{P}^{n-1}r_{A}H\Big(N_{sur} - N_{bulk}\Big).
\end{equation}
The respective degrees of freedom can be obtained 
from equations (\ref{deFms}), by replacing Hubble parameter with $r_A^{-1}.$ On substituting the variations of the entropy and horizon volume, the first law can be expressed as,
\begin{equation}\label{7eqn:llg}
\begin{split}
n\Omega_{n}r_{A}^{n+1}\dot{r}_{A} \sum_{i=1}^{m}ic_{i}r_{A}^{-2i}=r_{A}HL_{P}^{n-1}\Bigg[n\Omega_{n}r_{A}^{n+1} \sum_{i=1}^{m}\frac{ c_{i}r_{A}^{-2i}}{L_{P}^{n-1}} + \Bigg.  \\  \Bigg. \frac{\Omega_{n}r_{A}^{n+1} 8\pi ((n-2)\rho + np) }{(n-1)}\Bigg].
\end{split}
\end{equation}
As in the previous section, we now have to identify rate of change effective volume and have to use the respective Friedmann equation,
\begin{equation}\label{7eqn:lovelockFried}
\sum_{i=1}^{m}i\hat{c_{i}}\Bigg(H^{2}+\frac{k}{a^{2}}\Bigg)^{i}=\frac{16\pi L_{P}^{n-1}}{n(n-1)}\rho.
\sum_{i=1}^{m}i\hat{c_{i}}r_A^{-2i}=\frac{16\pi L_{P}^{n-1}}{n(n-1)}\rho.
\end{equation}   
Having this, it is easy to show that, equation (\ref{7eqn:llg}), will reduces the law of emergence given in equation (\ref{7eqn:sh}). 

In this section, we have derived the law of emergence proposed by Padmanabhan and its generalizations from the unified first law in the context of Einstein, 
Gauss-Bonnet and Lovelock gravities. This result emphasizes the fact that even though the law of emergence gets modified and generalized to take different forms, 
all these generalizations could be derived from the thermodynamic identity $TdS=dE+WdV,$ which has the same form regardless of the gravity theories. 
Although the basic assumptions that have been used for formulating these different generalized versions are different, all these could be derived from the same thermodynamic identity, 
in Einstein, Gauss-Bonnet and Lovelock gravities. These results point to a strong link between the law of emergence and the unified first law. We can say that the first law 
of thermodynamics is the backbone of the law of emergence.

\section{Feasibility of formulating the Law of Emergence with proper invariant volume}
In the previous  analysis the areal volume has been used in formulating  
the expansion law \cite{sheykhi} in non-flat universe, which 
resembles the volume of a sphere in Euclidean space \cite{HKT}. 
In fact areal volume can only account for the volume of space inside the apparent horizon when curvature parameter, $k=0$. 
On the other hand, the proper invariant volume surrounded by the apparent horizon is given as 
\begin{equation}\label{involume}
V_k=4\pi a^3\int_{0}^{(r_A/a)}dr\frac{r^2}{\sqrt{1-kr^2}}.
\end{equation}
Eune and Kim \cite{modthanu} have extended the
law of emergence to a non-flat universe, 
by employing the proper invariant volume, which is given by, 
\begin{equation}\label{ 8Myungseok_expasnion_law}
\frac{dV_k}{dt}=(L_{P}^2)_k^{eff}
\big(N_{sur}-\epsilon N_{bulk}\big),
\end{equation}
where $(L_{P}^2)^{\text{eff}}=L_{P}^2f_k(t),$ is the effective Plank length, which is time dependent. Here the function $f_k(t)$ is defined to be
\begin{equation}\label{fkt}
f_k(t)=\frac{\bar{V}_k[\dot{r}_AH^{-1}/r_A+(r_A/H^{-1})(H^{-1}/r_A-V_k/\bar{V}_k)]}{V_k(\dot{r}_AH^{-1}/r_A+V_k/\bar{V}_k-1)},
\end{equation}
where $\bar{V}_k=\frac{4\pi r_A^3}{3}$ is the areal volume. The authors have derived the Friedmann equations from this.
But, this work has been criticised mainly due to redefinition of the fundamental constant as a time varying one.
\cite{modthanu_critic}

It is thus necessary to see the consequences, when one try formulate the law of emergence using
invariant volume by preserving the fundamental nature of Planck's constant.  
Considering equation (\ref{involume}), the time derivative of invariant volume can be expressed as, 
\begin{equation}
\frac{dV_k}{dt}=3HV_k-Ar_A^2\frac{\ddot{a}}{a},
\end{equation}
where $A=4\pi r_A^2$. Using the second Friedmann equation, $(\ddot{a}/a)=(-4\pi L_{P}^2/3)(\rho+3P),$ 
one can express the difference in the number of degrees of freedom as 
\begin{equation}
\Delta N=\frac{A}{L_{P}^2}-\frac{3V_k}{L_{P}^2}\frac{\ddot{a}}{a}.
\end{equation}
Using this the rate of change of the proper volume 
can be expressed as 
\begin{equation}\label{8expasnion_new}
\frac{dV_k}{dt}=\frac{\bar{V}_k}{V_k}\big(L_{P}^2\Delta N\big)+3V_kH-\frac{A\bar{V}_k}{V_k}.
\end{equation}
This is entirely different from the law of emergence proposed by Padmanabhan, $dV/dt = L_{P}^2 (N_{surf}-N_{bulk}).$ 
This indicate that 
the expansion cannot be formulated appropriately,
by employing the proper invariant volume instead of areal volume. The authors of 
\cite{modthanu_critic} made a similar conclusion 
that
it is impossible to formulate the law of emergence in its original form
unless one defines a complicated time-dependent Planck length for a non-flat universe. 

 We have already shown that, 
the first law of thermodynamics can be considered as the backbone of the expansion law. Hence 
the failure in formulating the expansion law using invariant volume 
indicates that the issue is much deeper, such that the first law of thermodynamics may face similar problem when formulate it by  using invariant volume. Therefore let us now check the status of the first law of thermodynamics with proper invariant volume for a non-flat universe \cite{HKT}.

Now let us consider the unified first law in the $TdS=dE-WdV.$ 
Using Bekenstein's entropy relation, $S=A/4L_{P}^{2}$ and by taking the temperature of the horizon as $T=(-1/2\pi r_A)(1-(\dot{r}_A/2Hr_A)$ 
the left hand side of the previous equation  can be expressed as,
\begin{equation}\label{8lhs}
TdS = \frac{-dr_A}{L_{P}^{2}}\biggr(1-\frac{\dot{r}_A}{2Hr_A}\biggr),
\end{equation}   
which is independent of the choice of the volume. On the other hand, the  R.H.S.  can be obtained, by using  equation (\ref{8expasnion_new}), and the relation $E=\rho V_k$ as,
\begin{equation}\label{8rhs1}
dE-WdV_k = V_kd\rho +\frac{3}{2}(\rho+p)Hdt \biggr(V_k-\frac{4\pi r_A^2}{3H}+\frac{4\pi r_A \dot{r}_A}{3H^2}\biggr).
\end{equation}
This expression can be further simplified, with the help of continuity 
equation, 
$d\rho= -3H(\rho+p)dt,$ and the first Friedmann equation as, 
\begin{equation}\label{8rhs3}
dE-WdV_k=-\frac{dr_A}{L_{P}^2}\biggr(\frac{V_k}{2\bar{V}_k}-\frac{1}{2Hr_A}\biggr)+\frac{TdS}{H r_A},
\end{equation}
where $\bar{V}_k=4\pi/3H^3,$ the volume of the horizon of the corresponding flat universe.
Comparing this equation with equation(\ref{8lhs}), one can see that, 
the terms $dE-WdV_k,$  and  $TdS,$ differ by
$\displaystyle \left(\frac{V_k}{2\bar{V}_k}-\frac{1}{2Hr_A}\right).$ This difference proves that it is not possible to 
formulate the first law of thermodynamics, with the proper invariant volume in a non-flat universe. But, for a flat universe, where $V_k =\bar{V}_k$ and $H=r_A^{-1},$ the additional term vanishes and $dE-WdV_k$ will reduce to $TdS.$ 

We will now check, whether it is possible to regain the first law in a proper form for a non-flat universe by treating the Plank length as a function of cosmic time as in reference\cite{modthanu}. 
By using such a time dependent Planck length, $(L_{P}^2)^{\text{eff}}=L_{P}^2f_k(t)$ with $f_k(t)$   
as in equation (\ref{fkt}), the $TdS$ term will be modified as 
\begin{equation}\label{8TdS_modfied}
TdS=\frac{-1}{2L_{P}^{2}}\biggr(1-\frac{\dot{r}_A}{2Hr_A}\biggr)\bigg(\frac{2f_kdr_A-r_Adf_k}{f_k^2}\bigg).
\end{equation}
Proceeding as previously the R. H. S. of the first law takes the form
\begin{equation}\label{8rhs_modifed}
\begin{split}
dE-WdV_k=\frac{TdS}{H r_A}-\biggr(1-\frac{\dot{r}_A}{2Hr_A}\biggr)\frac{df_k}{L_{P}^2f_k^2H}- \\ \frac{1}{2L_{P}^2}\biggr(\frac{V_k}{2\bar{V}_k}-\frac{1}{2Hr_A}\biggr)\bigg(\frac{2f_kdr_A+r_Adf_k}{f_k^2}\bigg).
\end{split} 
\end{equation}
According to this equation, $dE-WdV_k$ 
does not reduce to $TdS$ with proper invariant volume.
Apart from 
a term, proportional to $\displaystyle \left(\frac{V_k}{2\bar{V}_k}-\frac{1}{2Hr_A}\right),$ like in equation (\ref{8rhs3}), there also arise an 
additional 
term proportional to $df_k/L_{P}^2H,$  which makes $dE-WdV_k$ different from $TdS.$ However for a flat universe for which $r_A=H^{-1},$ these extra terms will vanish since $f(t)_{k=0}=1$ and ${df_{k=0}}=0$, and hence the R. H. S. of equation (\ref{8rhs_modifed}) will reduces to $TdS$. Our results show that, 
even the use of 
time dependent Planck length, as proposed in \cite{modthanu}, 
cannot safeguard the first law of thermodynamics with proper invariant volume in a non-flat universe. Thus, it is impossible to formulate the unified first law of thermodynamics with the use of invariant volume in non-flat universe \cite{HKT}.

In this section, we have analysed the feasibility of formulating the law of emergence and the first law of thermodynamics in a non-flat universe. It is shown that both the law of emergence and the first law of thermodynamics demand the use of areal volume instead of proper invariant volume.
We have provided a more profound result than resolving the uncertainty
in the choice of volume used for formulating the law of emergence. It is shown that the unified first law can not be formulated at the apparent horizon with the use of proper invariant volume. The failure of the unified first law of thermodynamics, when one uses the proper invariant volume is a new result in the context of
cosmology. This provides justification for the conventional use of the areal volume of the apparent horizon.
 Moreover, it also
indicates that the invariant volume will not be a good choice to describe not only the law of emergence but
any thermodynamic process in the cosmological context.
 
 We have seen that the law of emergence could be derived from the first law thermodynamics. But, we can not say it is equivalent to first law. It is to be noted with much importance that, the law of emergence contain an upper bound on $N_{bulk},$ such that 
$N_{bulk} $ should not exceed the $N_{surf}.$ Curiously, such a bound on $N_{bulk}$ is not being taken care of by the first law of thermodynamics. In the following section we will analyse this matter in detail and 
will establish that, it is strongly coupled to the maximisation of the horizon entropy. 
\section{Holographic equipartition and the maximization of horizon entropy}
As is well known, every macroscopic system evolves to a state of thermodynamic equilibrium consistent with their constraints \cite{Callen1}.
The entropy of such systems
should attain a certain maximum value in the long run.
Pavon and Radicella\cite{Diego1} have shown that a Friedmann universe with a Hubble expansion history can behave as an ordinary macroscopic system 
with the horizon entropy tending to some maximum value. As we have discussed above, it is possible to describe  the evolution of the
universe as being driven by the departure from the holographic equipartition, and the universe is proceeding to a state that
satisfies the holographic equipartition. In this context, it is of interest to analyse the connection between holographic equipartition and horizon entropy maximization. 
Now, we study the problem whether the holographic equipartition explicitly
implies the maximization of horizon entropy or not in the context of Einstein, Gauss-Bonnet and Lovelock gravities \cite{KT,KT2}.

\subsection{Maximization of horizon entropy}
\label{sec:chp5_2}
In \cite{Diego1}, Pavan and Radicella have shown that our universe behaves as an ordinary macroscopic system that proceeds to a state of maximum entropy consistent with the two constraints, 
\begin{equation}\label{eqn:conditions1}
\dot S \geq 0, \, \textrm{always};  \, \, \, \, \, \,  \ddot S <0, \, \, \textrm{at \, least \, in \, long \, run.}
\end{equation}
Now, we will extend this procedure to  $n+1$ dimensional Einstein's
gravity, Gauss-Bonnet gravity
and Lovelock gravity for a spatially flat universe.
The rate of
change of horizon entropy with the progress of cosmic time can be obtained from equation (\ref{entropy1}) as,
\begin{equation}\label{5eqn:dns}
\dot S = -\frac{n(n-1)\Omega_n }{4L_{P}^{n-1}} \frac { \dot H}{H^n}.
\end{equation}
If the Hubble parameter $H$ is assumed to be always positive for an expanding universe, the horizon entropy will not decrease, if $\dot H\leq0$.  Now, we will check whether this horizon entropy is getting maximized in the long run, by analyzing the behaviour of $\ddot S$.

From equation (\ref{5eqn:dns}), one can easily find $\ddot S$ as,
\begin{equation}\label{5eqn:ddns}
\ddot S = \frac{n(n-1)\Omega_n }{4L_{P}^{n-1}} \left[ \left(\frac{n {\dot H}^2}{H^{n+1}}\right) - \left(\frac{\ddot H}{H^n} \right) \right].
\end{equation}
Then, we can immediately find the constraint for horizon entropy maximization,
\begin{equation} \label{5eqn:ineq2}
\left[n \left(\frac{\dot H^2}{H^{n+1}}\right) - \left(\frac{\ddot H}{H^n} \right)\right] < 0,
\end{equation}
in the asymptotic limit. If $\ddot{H}>0$ and $\dot{H} \to 0 $ in the asymptotic limit, the above inequality holds true for an expanding universe. We will now try to extend these results to Gauss-Bonnet and Lovelock gravity.

In Gauss-Bonnet gravity,  the rate of change of entropy can be calculated from equation (\ref{7equ:S}) as,
\begin{equation}\label{5eqn:gbdns}
\dot S = -\frac{n(n-1)\Omega_n }{4L_{P}^{n-1} H^n}  (1+{2\tilde\alpha H^2})\dot H ,
\end{equation}
where $n\geq 4$.
Then the generalized second law will be satisfied if,
\begin{equation}
(1+{2\tilde\alpha H^2})\dot H \leq 0.
\end{equation}
Taking the time derivative of equation (\ref{5eqn:gbdns}), we find
\begin{equation}\label{5eqn:gbddns}
\ddot S = \frac{n(n-1)\Omega_n }{4L_{P}^{n-1} H^{n+1}}  \left[{\dot H}^{2}[{n+(2n-4)\tilde\alpha H ^{2}}] - 
[{H}{\ddot H}(1+{2\tilde\alpha \ H^{2})}] \right].
\end{equation}
The
entropy maximization demands,
\begin{equation}\label{5eqn:con3}
\left[ {\dot H}^{2} \left[ {n+(2n-4)\tilde\alpha H ^{2}} \right]  -  {H}{\ddot H}(1+{2\tilde\alpha \ H^{2})}\right]  < 0
\end{equation}
in the final stage. 

We will now move to the more general Lovelock gravity. The rate of change of horizon entropy in Lovelock gravity can be obtained from equation
(\ref{7entropy}) as,
\begin{equation}\label{5eqn:llds}
\dot S= -\frac{n(n-1)\Omega_n {\dot H}}{4L_{P}^{n-1} H^{n+2}} \sum_{i=1}^m {i} \hat{c_i}{H}^{2i} .
\end{equation}

The generalized second law will be satisfied if
\begin{equation}
\dot H\sum_{i=1}^m {i} \hat{c_i}{H}^{2i} \leq 0.
\end{equation}
Now we will consider the second derivative of entropy by differentiating equation (\ref{5eqn:llds}),
\begin{equation}\label{5eqn:lldds}
\ddot S= \frac{n(n-1)\Omega_n }{4L_{P}^{n-1}} \sum_{i=1}^m i \hat{c_i} H ^{2i} \left[ {(n+2-2i) \frac{{\dot H}^2}{H^{n+3}}} -\frac{\ddot H}{H^{n+2}} \right] .
\end{equation}

The horizon entropy tends to some
maximum value if,
\begin{equation}\label{5eqn:con4}
\sum_{i=1}^m i \hat{c_i} H ^{2i} \left[ {(n+2-2i) \frac{{\dot H}^2}{H^{n+3}}} -\frac{\ddot H}{H^{n+2}} \right]<0
\end{equation}
in the long run. 

Here we have obtained the constraints imposed by the horizon entropy maximization in the context of  
Einstein, Gauss-Bonnet and Lovelock
gravities. In the upcoming section we shall see whether the law of emergence guarantees the maximization of horizon entropy \cite{KT,KT2}.

\subsection{Holographic equipartition vs entropy maximization : Analysis on Padmanabhan's Proposal}

Here, we  show that the law of emergence suggested by Padmanabhan effectively implies the maximization of horizon entropy.
 Using equations (\ref{eqn:dVdt1}) and (\ref{entropy1}), the rate of change of horizon entropy in Einstein gravity could be expressed in terms of the degrees of freedom as, 

\begin{equation} \label{eqn:sdot2}
\dot S = \frac{2\pi}{H} \left(1-\epsilon \frac{N_{bulk}}{N_{surf}} \right).
\end{equation}
For $\dot S \geq 0,$ the condition to be satisfied is
\begin{equation}
1-\epsilon \frac{N_{bulk}}{N_{surf}} \geq 0 ,
\end{equation}
which is equivalent to $(N_{surf}-\epsilon N_{bulk}) \geq 0$. Since the bulk degrees of freedom  can not exceed the surface degrees of freedom according to Padmanabhan's proposal, the above inequality is always satisfied which in turn guarantees the non negativity of $\dot{S}$.

Differentiating equation (\ref{eqn:sdot2}) with respect to cosmic time, one gets,
\begin{equation}
\ddot S = -\frac{2\pi\dot H}{H^2} \left( 1- \epsilon \frac{N_{bulk}}{N_{surf}} \right) +
\frac{2\pi}{H} \frac{d}{dt} \left(1-\epsilon \frac{N_{bulk}}{N_{surf}} \right).
\end{equation}
From equation (\ref{eqn:sdot2})  we can easily write '$-\dot H /H^2 = 1-\epsilon \frac{N_{bulk}}{N_{surf}}$ ', and the above equation could be simplified as 
\begin{equation}
\ddot S = 2\pi \left( 1- \epsilon \frac{N_{bulk}}{N_{surf}} \right)^2 +
\frac{2\pi}{H} \frac{d}{dt} \left(1-\epsilon \frac{N_{bulk}}{N_{surf}} \right).
\end{equation}

Since the bulk degrees of freedom will be equal to the  the surface degrees of freedom in the final stage, the first term in the above equation vanishes in the long run.  As the universe is trying to decrease the holographic discrepancy, $ \frac{d}{dt} \left(1-\epsilon \frac{N_{bulk}}{N_{surf}} \right)$ must be negative. This together guarantees the condition, $\ddot S < 0$ in the long run, which ensures the maximization of horizon entropy \cite{KT,KT2}.

Next, our aim is to see whether the generalized holographic equipartition in \cite{Cai1} and \cite{Yang} leads to the maximization of entropy.
\subsection{Holographic equipartition and entropy maximization: Analysis of Cai's proposal}
We shall now consider the generalized holographic equipartition suggested by Cai and will see whether it leads to the maximization of horizon entropy. In (n+1) dimensional Einstein's gravity,  the time derivative of the cosmic volume,
$V=\Omega_n /{H}^{n}$ can be obtained as,
\begin{equation}
{dV\over dt}= -n\Omega_n \frac{\dot H}{H^{n+1}}.
\end{equation}
Now using equation (\ref{5eqn:dns}), we get
\begin{equation}\label{5volume change in terms entropy1}
{dV\over dt} = \frac{4L_{P}^{n-1} }{H(n-1)}  \dot S.
\end{equation}
Then, the law of emergence in equation (\ref{emergndimen}), can be written as,
\begin{equation}\label{5Eeqncon1}
\dot S ={\frac{(n-2)H}{2}(N_{surf}- N_{bulk})}.
\end{equation}
For an expanding universe which satisfies the holographic equipartition, we have 
${N_{surf}- N_{bulk}}\geq0$, which in turn ensures the non negativity of $\dot S$.

Differentiating the above equation, we get
\begin{equation}\label{5Eeqncon11}
\ddot S ={\frac{(n-2)\dot H}{2}(N_{surf}- N_{bulk})}+{{\frac{(n-2)H}{2}}}  {{d\over dt}(N_{surf}- N_{bulk})} .
\end{equation}
If the universe satisfies holographic equipartition in the final state $N_{bulk}$ approaches $N_{surf}$ in the long run and the first term in the above expression vanishes. Also, since the universe is trying to decrease the holographic discrepancy, ${d\over dt}(N_{surf}- N_{bulk})$ will be negative. Hence $\ddot{S}$ will be negative in the final state, implying the maximization of horizon entropy.

We shall now investigate whether law of emergence implies horizon entropy maximization in the context of Gauss-Bonnet gravity. From equations
(\ref{7eqn:sh}) and (\ref{5eqn:gbdns}), we can write the law of emergence as,
\begin{equation}
	\dot S ={\frac{(n-2)H}{2}(N_{surf}- N_{bulk})}.
\end{equation}
where '$S$' is the horizon entropy in Gauss-Bonnet gravity. Even though the surface degrees of freedom takes a much complex form, here also we can assure the non negativity of $\dot{S}$, since ${N_{surf}- N_{bulk}}\geq0$, as the universe satisfies the holographic equipartition. As we have seen earlier, we can obtain the second derivative of horizon entropy as, 

\begin{equation}\label{5GBeqncon11}
\ddot S ={\frac{(n-2)\dot H}{2}(N_{surf}- N_{bulk})}+{{\frac{(n-2)H}{2}}}  {{d\over dt}(N_{surf}- N_{bulk})}.
\end{equation}

Here also, the first term will reduce to zero in the final state, since $N_{bulk}$ approaches $N_{surf}$. Since the universe is trying to decrease the holographic discrepancy, the above equation assures the negativity of $ \ddot S $ in the long run and thus implies the maximization of horizon entropy.

These results can be extended to Lovelock gravity \cite{KT2}.  One can prove that the law of emergence in equation (\ref{7eqn:lovelockexp}) leads to horizon entropy maximization following the same procedure.
We can see that the law of emergence suggested by Cai leads to the maximization of horizon entropy in the context of Einstein, Gauss-Bonnet and Lovelock gravities.

\subsection{Holographic equipartition and entropy maximization: Analysis of Yang et.al.'s proposal}
Here we will investigate whether the generalized version of law of emergence proposed in \cite{Yang} leads to the maximization of horizon entropy.
We shall now consider the law of emergence in equation (\ref{yangeq}), which is proposed in the context of Gauss-Bonnet Gravity. Combining the equations (\ref{5eqn:gbdns}), (\ref{yangeq}) and (\ref{eqn:yang1}), one can express the generalized law of emergence in Gauss-Bonnet gravity as,

\begin{equation}
\dot S=\frac{(n-1)H}{4}(1+2\tilde\alpha H^{2}) f(\Delta N,N_{surf}).
\end{equation}
 The horizon entropy will not decrease, if $f(\Delta N,N_{surf})$ is chosen in such a way that $\dot S \geq 0$. Taking the time derivative of the above equation, we get,

\begin{equation}
\begin{split}
\ddot S=\frac{(n-1)}{4}\frac{d}{dt} \left( H(1+2\tilde\alpha H^{2}) \right) f(\Delta N,N_{surf})+  \\
\frac{(n-1)}{4} H (1+2\tilde\alpha H^{2}) \frac{d}{dt} \left( f(\Delta N,N_{surf}) \right).
\end{split}
\end{equation}
The horizon entropy will tend to a finite maximum if the function $f(\Delta N,N_{surf})$ obeys the condition,
\begin{equation}
\frac{d}{dt} \left( f(\Delta N,N_{surf}) \right)<0
\end{equation}
in the final stage, provided the first term vanishes in the long run, which demands
\begin{equation}
\frac{d}{dt} \left( H(1+2\tilde\alpha H^{2}) \right) f(\Delta N,N_{surf}) =0
\end{equation}
in the asymptotic limit.

Now, we will move to Lovelock gravity. From equations (\ref{5eqn:llds}), (\ref{yangeq}) and (\ref{fdeltaN}), we can express the law of emergence suggested by Yang.et.al as,
\begin{equation}
\dot S=\frac{(n-1)}{4}\sum_{i=1}^m  i\hat{c_i} H^{2i-1} f(\Delta N,N_{surf}).
\end{equation}
The horizon entropy will never decrease, if $f(\Delta N,N_{surf})$ is chosen in such a way that $\dot S \geq 0$. Differentiating the above equation, we readily get,

\begin{equation}
\begin{split}
\ddot S=\frac{(n-1)}{4}\frac{d}{dt} \left( \sum_{i=1}^m  i\hat{c_i} H^{2i-1} \right) f(\Delta N,N_{surf})+  \\
\frac{(n-1)}{4}\sum_{i=1}^m  i\hat{c_i} H^{2i-1}
\frac{d}{dt} \left( f(\Delta N,N_{surf}) \right) .
\end{split}
\end{equation}
As we have seen earlier, the horizon entropy finally tends to a finite maximum, if the function $f(\Delta N,N_{surf})$ satisfies the inequality,
\begin{equation}
\frac{d}{dt} \left( f(\Delta N,N_{surf}) \right)<0
\end{equation}
in the final stage, provided the first term vanishes in the long run, which demands,
\begin{equation}
\frac{d}{dt} \left( H(1+2\tilde\alpha H^{2}) \right) f(\Delta N,N_{surf}) =0
\end{equation}
in the final stage.
We have found the constraints imposed by the horizon entropy maximization on the function $f(\Delta N,N_{surf})$. If the generalized version of the law emergence obeys the above mentioned constraints the horizon entropy of the universe will get maximized. Here we have discussed the connection between the generalized law of emergence and the horizon entropy maximization for a spatially flat universe in the context of Einstein, Gauss-Bonnet and Lovelock gravities. One can also extend these results to the non flat universe \cite{KTarxive}. 

In this section,
we have shown that the law of emergence proposed by Padmanabhan effectively implies the maximization of horizon entropy. In other words, the tendency for decreasing the holographic discrepancy or the tendency for equalizing the degrees of freedom can be interpreted as a tendency for maximizing the horizon entropy. We have investigated the consistency of the law of emergence with the horizon entropy maximization in the context of Einstein, Gauss-Bonnet,
and Lovelock gravities. First, we extended the procedure suggested by Pavan and Radicella in the context of Einstein gravity, to Gauss-Bonnet and more general Lovelock gravities and obtained the constraints of horizon entropy maximization in each gravity theory. We then analysed the consistency of the two generalized versions of the law of emergence suggested by  Cai and Yang et. al. with the  horizon entropy maximization. 
These results point at the strong connection between the law of emergence and the horizon entropy maximization for a spatially flat universe beyond Einstein's gravity. It is also possible to extend these results to a non-flat universe following the same procedure \cite{KTarxive}.

\section{Conclusions} 
The first theme of this review is to provide a rapid overview on the evolving perspectives on gravity and spacetime, which could be summarised as follows.

 Despite its great success, we can not say general relativity is complete or flaw-less especially due to the existence of singularities. Even though the scientific community hope to resolve the singularity problem by formulating a quantum theory of gravity, they have not yet succeeded. At the same time several higher dimensional gravity theories have also been suggested. Meanwhile, a curious connection between gravity and thermodynamics have been established, which does not have a natural explanation in the framework of general relativity. This paved the way for Padmanabhan's emergent gravity paradigm, where gravity is considered as an emergent phenomenon like elasticity, viscosity etc. Based on this emergent paradigm, Padmanabhan interpreted the cosmological evolution as the emergence of cosmic space with the progress of cosmic time.

The next theme of this review is to establish the relationship between the law of emergence and the principles of thermodynamics. Below we summarize the major findings, which establish the strong connection of the law of emergence with the unified first law of thermodynamics and the maximization of horizon entropy.

\begin{itemize}
	\item The law of emergence and its generalizations could be derived from the thermodynamic identity $TdS=dE+WdV,$ which is dubbed as the unified first law, in the context of Einstein, Gauss-Bonnet and Lovelock gravities.
	\item It is important to note that various generalizations of the law of emergence could be derived from the thermodynamical identity, that has the same form regardless of the 
	gravity theories.
	\item Both the law of emergence and the first law of thermodynamics necessitates the use of areal volume instead of proper invariant volume in a non-flat universe.
	\item The law of emergence effectively implies the maximization of horizon entropy in the context of Einstein, Gauss-Bonnet and Lovelock gravities. 

	\item The horizon entropy maximization impose an upper bound to the bulk degrees of freedom such that it should not exceed the surface degrees of freedom. First law of thermodynamics with this additional constraints imposed by horizon entropy maximization together leads to the law emergence.
\end{itemize}

When one asks, what this law of emergence indicates, we can say that it assures the maximization of horizon entropy. And as to its origin, we can say that it could be derived from the first law of thermodynamics.
Here, we have found satisfactory answers to the two
questions: what the law of emergence implies and what forms its basis from a thermodynamic perspective. While the law of emergence implies the maximization of 
horizon entropy its backbone is formed by the unified first law. It is also important to note that these results holds in Einstein, Gauss-Bonnet and Lovelock 
gravities for a universe with any spatial curvature. This gives us a new insight that the law of emergence is directly connected to the thermodynamics of horizons. Here we 
reach an astonishing conclusion that the entire thermodynamics is encrypted in the law of emergence.

The above results offer strong support to the law of emergence from a thermodynamic point of view. These results also point to some new possibilities for 
future research. 

\begin{itemize}
   	\item It is possible to arrive at the law of emergence from the first law of thermodynamics in a variety of gravity theories.
	\item It can be shown that the law of emergence leads to the maximization of horizon entropy in various theories of gravity.
	\item The validity of holographic equipartition or horizon entropy maximization can be used to impose constraints on the modified gravity parameters like the Gauss-Bonnet coefficient or the various coefficients in Lovelock gravity.	
	\item A generic method  can be formulated to derive the law of emergence from first law of thermodynamics, suitably fixing surface degrees of freedom and the form of total energy in the bulk. The same procedure can be used to find the possible form of the law of emergence in non-equilibrium situations; where we should start from the first law of thermodynamics in non-equilibrium description.
	\item One can also analyse the feasibility of describing a variety of cosmological 
	models using the law of emergence.
\end{itemize}
  We can also explore whether the more general evolution equations of the emergent gravity paradigm are linked with the thermodynamics of horizons. \\[.2in]
  
  \noindent \textbf{Acknowledgement}:  We thank Mahith M and Hareesh T for discussions. Hassan Basari V T acknowledges Cochin University of Science and Technology for research fellowship.\\[0.2in]
  
  \noindent \textbf{ Data availability }\\
  We haven't used any data for any of the analysis in this article.

\end{document}